\documentclass[%
 reprint,
 amsmath,amssymb,
 aps,
]{revtex4-2}
\usepackage{xcolor, soul}
\sethlcolor{yellow}
\usepackage{graphicx}
\usepackage{dcolumn}
\usepackage{bm}
\usepackage{lipsum}

\usepackage{subfigure}

\begin{document}


\title{Experimental investigation of the effect of topological insulator on the magnetization dynamics of ferromagnetic metal: $BiSbTe_{1.5}Se_{1.5}$ and $Ni_{80}Fe_{20}$ heterostructure}

\author{Sayani Pal, Soumik Aon, Subhadip Manna, Sambhu G Nath, Kanav Sharma $\&$ Chiranjib Mitra}
\email{Corresponding author:chiranjib@iiserkol.ac.in}
\affiliation{%
 Indian Institute of Science Education and Research Kolkata,\\
 Mohanpur 741246, West Bengal, India\\
 }

\date{\today}

\begin{abstract}
We have studied the spin-pumping phenomenon in ferromagnetic metal($Ni_{80}Fe_{20}$)/topological insulator($BiSbTe_{1.5}Se_{1.5}$) bilayer system to understand magnetization dynamics of ferromagnetic metal (FM) in contact with a topological insulator (TI). TIs embody a spin-momentum-locked surface state that spans the bulk band gap. Due to this special spin texture of the topological surface state, the spin-charge interconversion efficiency of TI is even higher than that of heavy metals. We evaluated the parameters like effective damping coefficient ($\alpha_{eff}$), spin-mixing conductance ($g_{eff}^{\uparrow \downarrow}$) and spin current density ($j_S^0$) to demonstrate an efficient spin transfer in $Ni_{80}Fe_{20}/BiSbTe_{1.5}Se_{1.5}$ heterostructure. To probe the effect of the topological surface state, a systematic low-temperature study is crucial as the surface state of TI dominates at lower temperatures. The exponential increase of $\Delta H$ for all different thickness combinations of FM/TI bilayers and the enhancement of effective damping coefficient ($\alpha_{eff}$) with lowering temperature confirms that the spin chemical potential bias generated from spin-pumping induces spin current into the TI surface state. Furthermore, low-temperature measurements of effective magnetization ($4\pi M_{eff}$) and magnetic anisotropy field ($H_k$) showed anomaly around the same temperature region where the resistivity of TI starts showing metallic behavior due to the dominance of conducting TI surface state. The anomaly in $H_k$ can result from the emerging exchange coupling between the TI surface state and the local moments of the FM layer at the interface without any long-range ferromagnetic order in TI at the interface.

\end{abstract}

\pacs{Valid PACS appear here}
\maketitle


\section*{Introduction}
Spintronics is one of the emerging fields that has witnessed remarkable progress on both fundamental and technological fronts over the past couple of decades. Phenomena like spin-orbit torque \cite{I2}, spin Hall effect\cite{I3}, giant magnetoresistance \cite{I4}, tunnelling magnetoresistance \cite{I5}, domain wall motion \cite{I6} provide basics for applications in memory devices\cite{I7}, storage technology\cite{I8}, logic gates \cite{I9} and magnetic sensors \cite{I10}. These devices utilize the spin degrees of freedom of electrons and their interaction with orbital moments through spin-orbit coupling. Complete knowledge of the process of generation, manipulation, and detection of spin degrees of freedom or the spin current is essential for widespread applications in this field. If one focuses on the currently available spin current generation processes, spin-pumping \cite{I13, I14} is one of the most efficient methods where the precessing magnetization in the ferromagnet (FM) injects spin current into the adjacent layer by transferring spin angular momentum. This raises a need to study the effect of spin pumping with special emphasis on exploring new materials which can give rise to significant spin-charge interconversion efficiency. Topological insulators (TI) are a new class of materials that have an interesting spin texture of the surface state, owing to spin-momentum locking\cite{I20, I21, I21a, I22}. The momentum direction of the electron in the surface state of TI is perpendicularly locked to its spin polarization direction. Thus the spin-charge interconversion for TI is even higher than the heavy metals which makes TIs suitable for spintronics application\cite{I17}. As surface states are robust against deposition of FM layers on top of TI \cite{Sup1}, the topological surface states remain intact and gapless \cite{Sup2}. TI/FM bilayers have been successfully used for the spin current generation in spin-pumping experiments\cite {3a,3b,FM/TI1, FM/TI2, Sup3}. The effect of spin pumping can be witnessed in the enhanced damping coefficient ($\alpha_{eff}$) value of the ferromagnet upon excitations of ferromagnetic resonance (FMR) because, in the spin pumping process, the net transfer of spin angular momentum into TI layer brings about an additional damping torque on the precessing magnetization in the FM. It is difficult to fabricate a perfect TI thin film where the bulk state of TI is completely insulating. Thus for a complete understanding of the effect of TI surface state on FM magnetization dynamics, the low-temperature study is necessary where the surface states of TI dominate.\\
In this paper, we present the study of the spin-pumping phenomenon in ferromagnetic metal (FM)/ topological insulator (TI) bilayer system. We chose $Ni_{80}Fe_{20}$ as the FM layer and $BiSbTe_{1.5}Se_{1.5}$ as the TI layer. Currently, $BiSbTe_{1.5}Se_{1.5}$ is one of the best 3D TI materials in which bulk conduction in thin films is negligible even at room temperature and the dominance of surface state is very prominent at lower temperatures \cite{BSTS1, BSTS2, BSTS3}. In our low-temperature measurements, we have witnessed exponential enhancement of FMR linewidth ($\Delta H$) and effective damping coefficient ($\alpha_{eff}$) at lower temperatures. It supports the proposal of the spin chemical potential bias induced spin current injection into the surface state of TI given by Abdulahad\textit{et al.}\cite{TI1}. For further investigation of the effect of the TI surface state on the FM magnetization, we have also studied low-temperature variations of effective magnetization and anisotropy field. We calculated the interfacial magnetic anisotropy of the bilayer to be in-plane of the interface. At low temperatures, this magnetic anisotropy field shows a hump-like feature concomitant with the resistivity behavior of $BiSbTe_{1.5}Se_{1.5}$ with temperature. It predicts the existence of exchange coupling between the surface states of TI and the local moments of the FM layer which acts perpendicular to the TI/FM interface.  We have also evaluated the values of spin-transport parameters like spin-mixing conductance, $g_{eff}^{\uparrow \downarrow}$ and spin current density, $j_s^0$ at room temperature to ensure a successful spin injection into the TI layer from the FM layer.\\

\section*{SAMPLE PREPARATION and CHARACTERIZATION}
For this particular work, we have prepared different thickness combinations of topological insulator(TI)/ferromagnet(FM) bilayer heterostructure. $BiSbTe_{1.5}Se_{1.5}(BSTS)$ has been taken as the TI material and Permalloy($Ni_{80}Fe_{20}$) has been used as the ferromagnetic material. BSTS thin films were grown on silicon (Si 111) substrate using pulsed laser deposition(PLD) technique \cite{S1, S2}. The target material was prepared using $99.999\%$ pure Bi, Sb, Te, and Se in a 1:1:1.5:1.5 stoichiometric ratio. The films were deposited through ablation of the target by a KrF excimer laser (248 nm, 25 ns pulse width) at a low repetition rate of 1Hz and $1.2 J cm^{-2}$ laser fluence keeping the substrate temperature fixed at $250^{0} C$ and the chamber partial pressure at 0.5 mbar (base pressure $2 \times 10^{-5}$ mbar) with a continuous flow of Ar gas. After deposition, TI films were immediately transferred into the thermal evaporation chamber for the deposition of the FM layer. Commercially available $99.995\%$ pure permalloy ($Ni_{80}Fe_{20}$) pallets were used for deposition.  The Py film was deposited \cite{S3} on top of TI film at a rate of 1.2$\AA$ (crystal monitor: Inficon SQM 160) keeping the chamber pressure fixed at $1\times 10^{-6}$ torr (base pressure $1\times 10^{-7}$ torr). For the characterization of the films X-ray diffraction analysis (XRD), field emission scanning electron microscope (FE-SEM) imaging, and atomic force microscopy (AFM) facilities have been used. X-ray reflectometry technique has been used for thickness measurements here. For convenience we are defining the BSTS of different thicknesses as follows: 10nm BSTS as BSTS1, 21nm BSTS as BSTS2, 28nm BSTS as BSTS3, and 37nm BSTS as BSTS4.

\section*{RESULTS and DISCUSSION}
\begin{figure*}
\centering
\subfigure[]{\includegraphics[width=6.5cm,height=5cm]{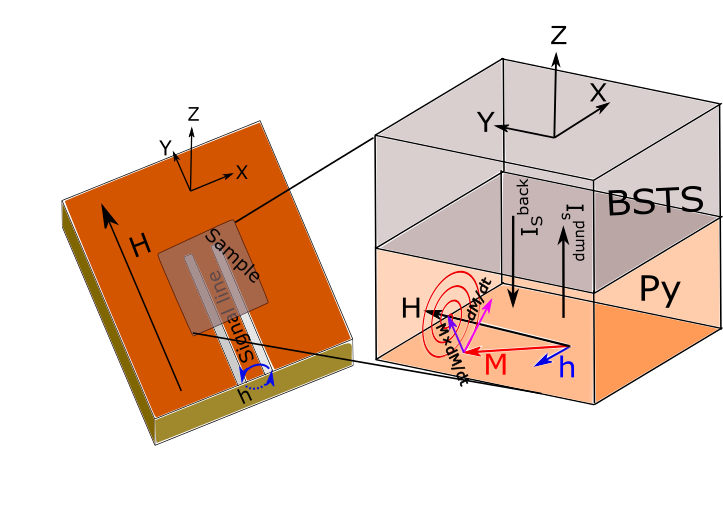}}
\subfigure[]{\includegraphics[width=6.5cm,height=5cm]{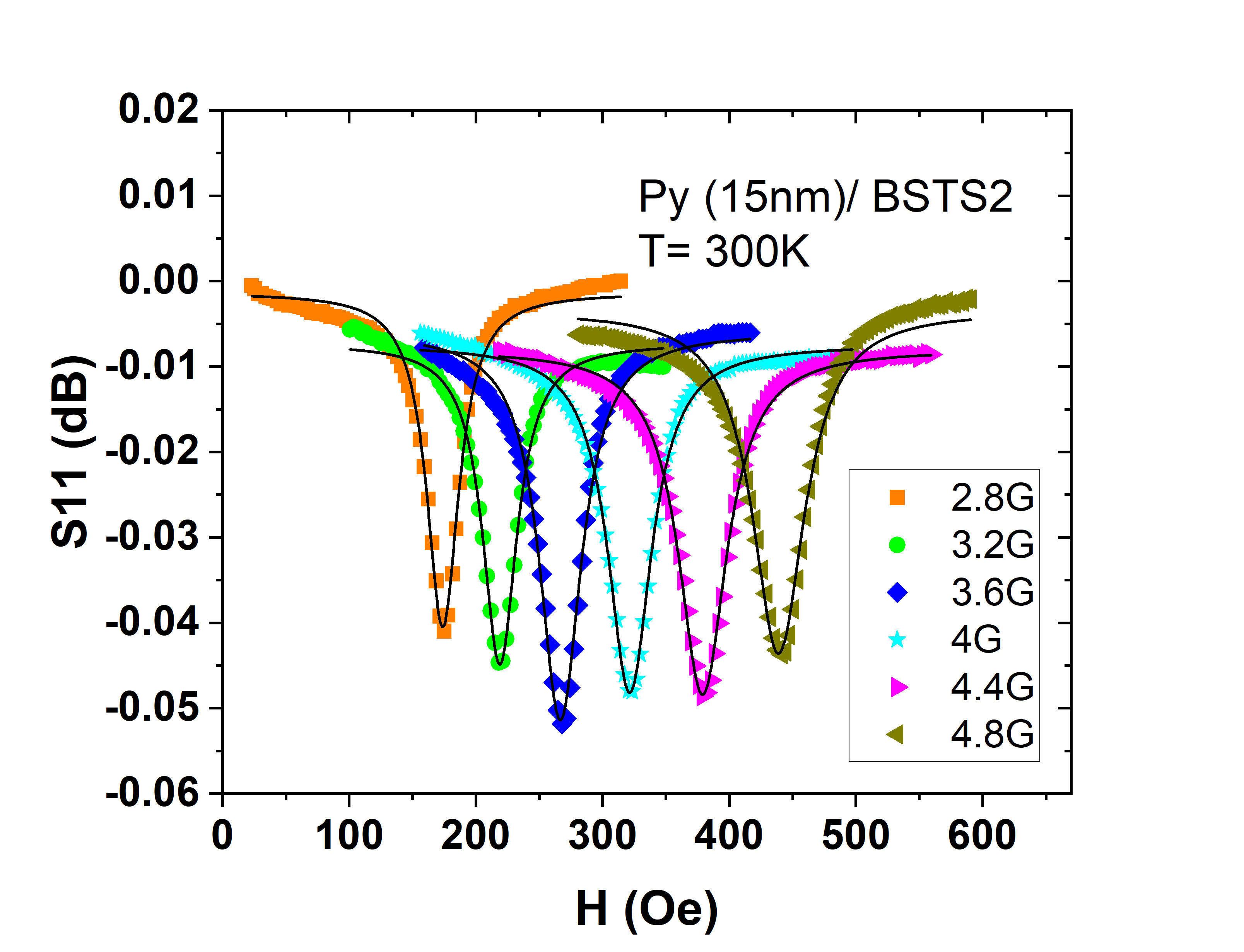}}
\caption{(a) In the left diagram, a schematic illustration of the experimental set-up has shown where the FM/TI bilayer is placed upside down on top of a CPW, and in the right diagram, net injected spin current ($I_S^{pump}$) due to spin-pumping into the TI layer (BSTS) from the FM layer (Py) has shown, it results faster magnetization relaxation in FM; (b) Ferromagnetic Resonance spectra of absorption at different frequencies for  Py/BSTS bilayer system at room temperature after background subtraction.}
\label{Fig1}
\end{figure*}
\begin{figure*}
\centering
\subfigure[]{\includegraphics[width=5.9cm,height=4.5cm]{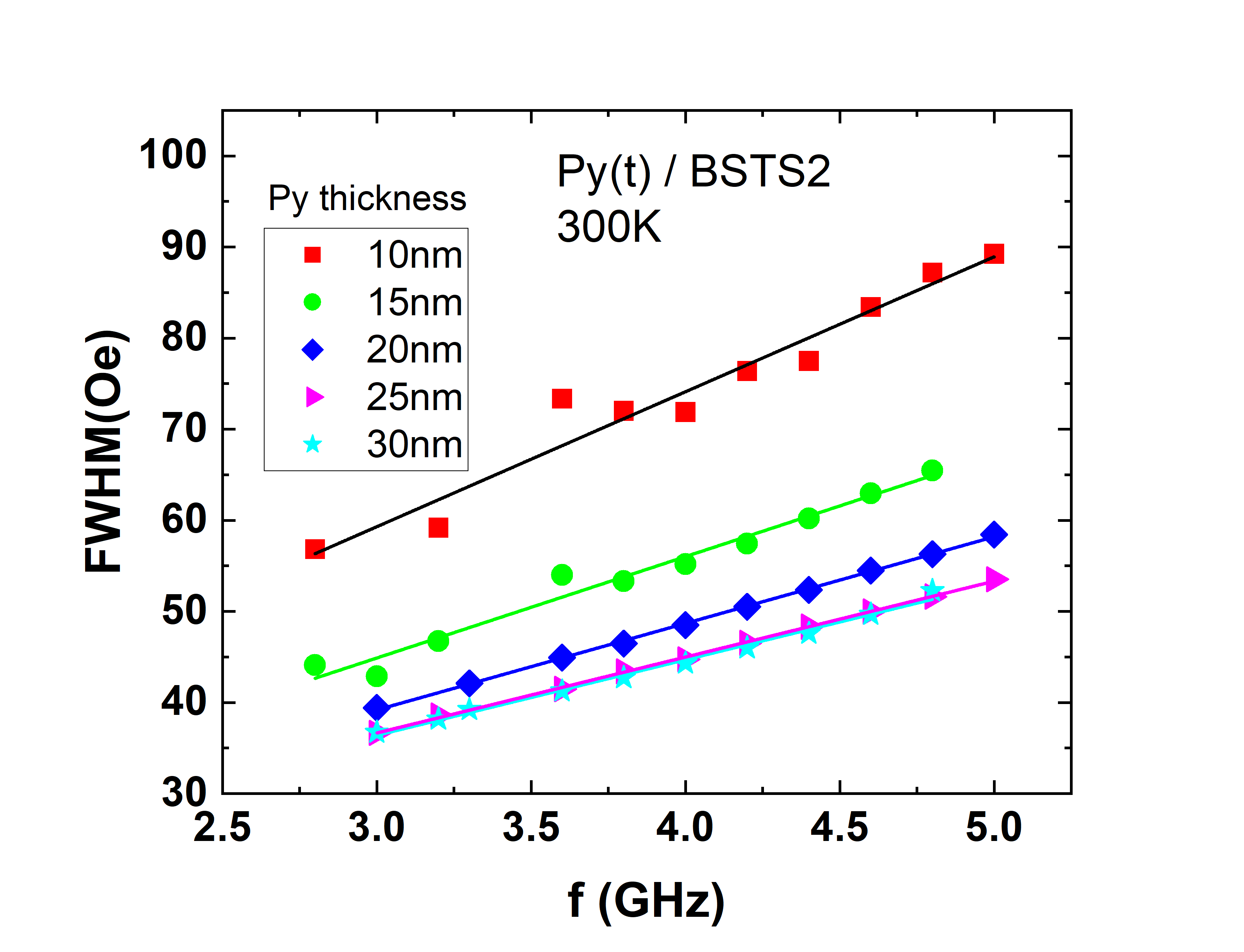}}
\subfigure[]{\includegraphics[width=5.9cm,height=4.5cm]{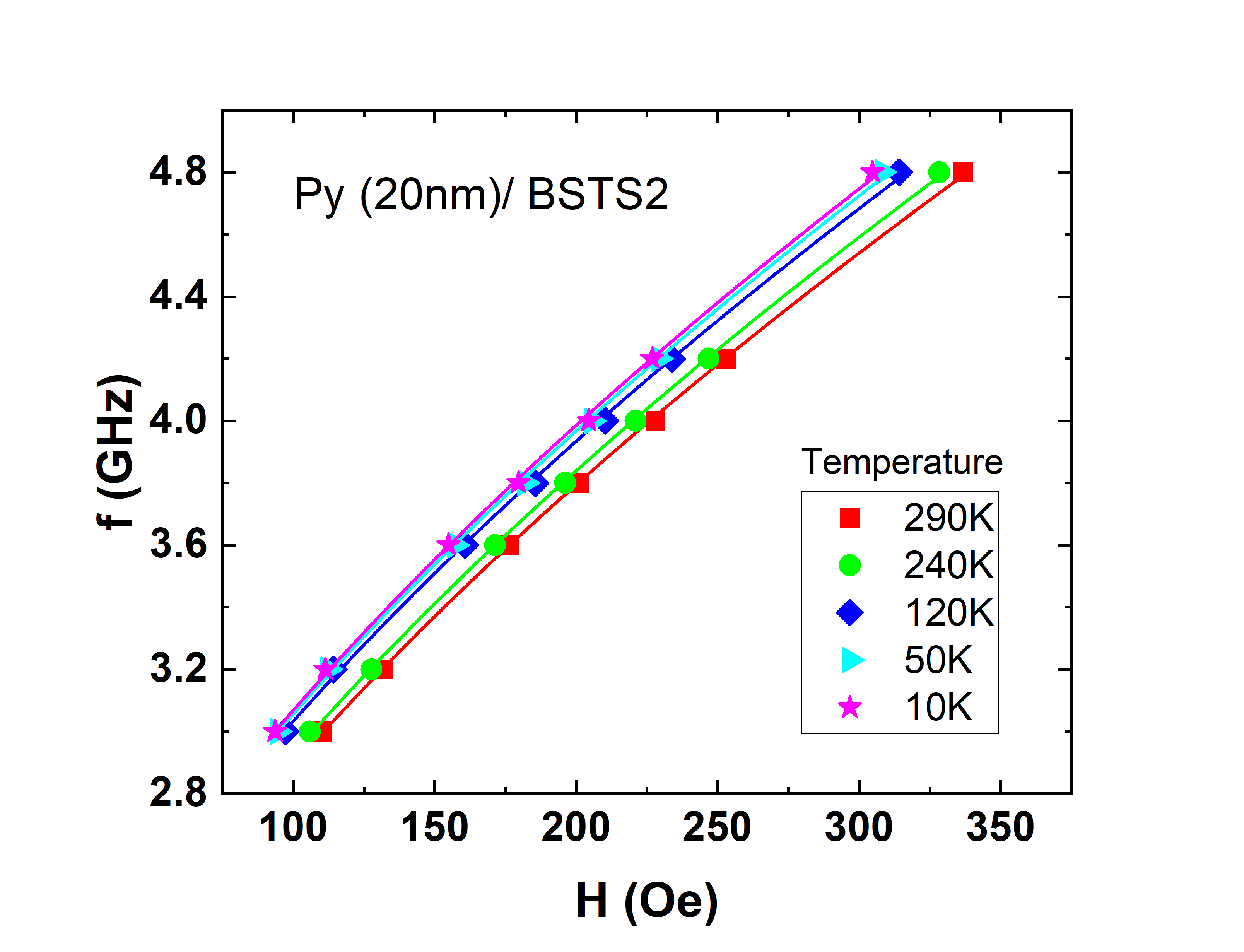}}
\subfigure[]{\includegraphics[width=5.9cm,height=4.5cm]{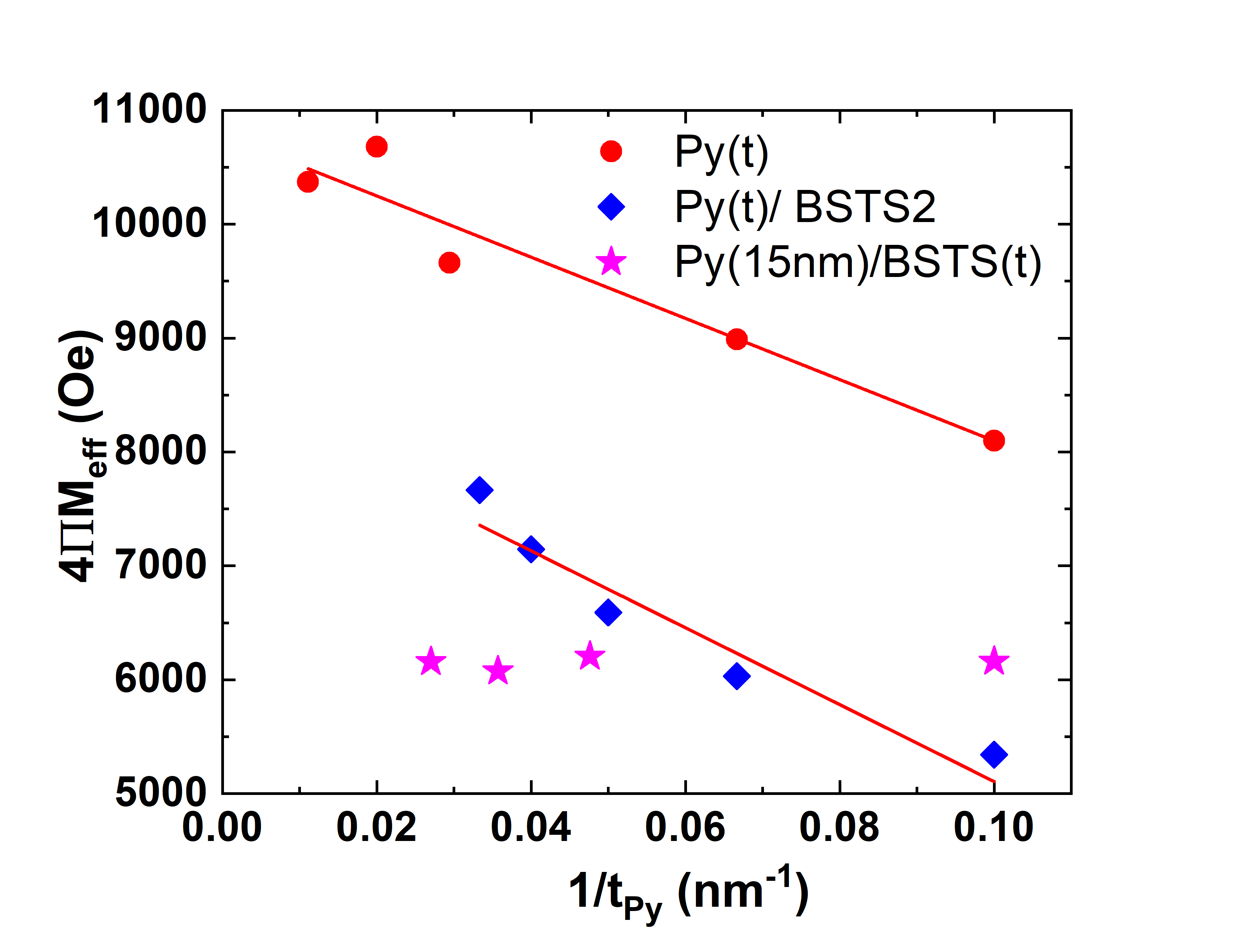}}
\caption{(a) Field linewidth ($\Delta H$) variation with resonance frequencies ($f$) at 300K for Py/BSTS bilayer samples with different Py thicknesses. Eq.\ref{Damping} has been used for fitting the curve and to determine the damping coefficient ;(b) Resonance field ($H$) vs. resonance frequency ($f$) for Py(20nm)/BSTS2 system at different temperatures . Eq.\ref{Kittel} has been used for fitting the curve and to determine the effective magnetization; (c) Effective magnetization ($4\pi M_{eff}$) variation with thickness of Py(t), Py(t)/BSTS2 and Py(15nm/BSTS(t) at room temperature. Eq\ref{Ms} has been used for fitting the curve and to evaluate saturation magnetization ($4\pi M_S$) and magnetic anisotropy field($H_k$).}
\label{Fig2}
\end{figure*}

For a systematic study of the FM/TI bilayer system, we have done in-plane FMR measurements in reflection mode geometry using a short-circuited CPW as shown in fig.\ref{Fig1}a. We obtained typical FMR signal at different microwave frequencies for Py(15nm)/BSTS2 sample in fig.\ref{Fig1}b. From the Lorentz formula fitting \cite{Lorentz} of the FMR signal we extracted the frequency dependence of the field linewidth ($\Delta H$ vs. $f$) and the resonance frequency vs. resonance field ($f$ vs $H$) data as shown in fig.\ref{Fig2}a and fig.\ref{Fig2}b respectively. These give us valuable information about the magnetization dynamics in ferromagnet which can be described within the framework proposed by Landau, Lifshitz, and Gilbert \cite{Gilbert},

\begin{equation}
\frac{d\vec{M}}{dt}=-\gamma \vec{M} \times \vec{H}_{eff} + \frac{\alpha_{eff}}{M_S} \vec{M} \times \frac{d \vec{M}}{dt}
\label{LLG}
\end{equation}
where, $\gamma$ is the gyromagnetic ratio, $\vec{M}$ is the magnetization vector, $M_S$ is the saturation magnetization, $H_{eff}$ is the effective magnetic field which includes the external field, demagnetization and crystalline anisotropy field and $\alpha_{eff}$ is the effective damping coefficient of the system.

\begin{figure*}
\centering
\subfigure[]{\includegraphics[width=7cm,height=5cm]{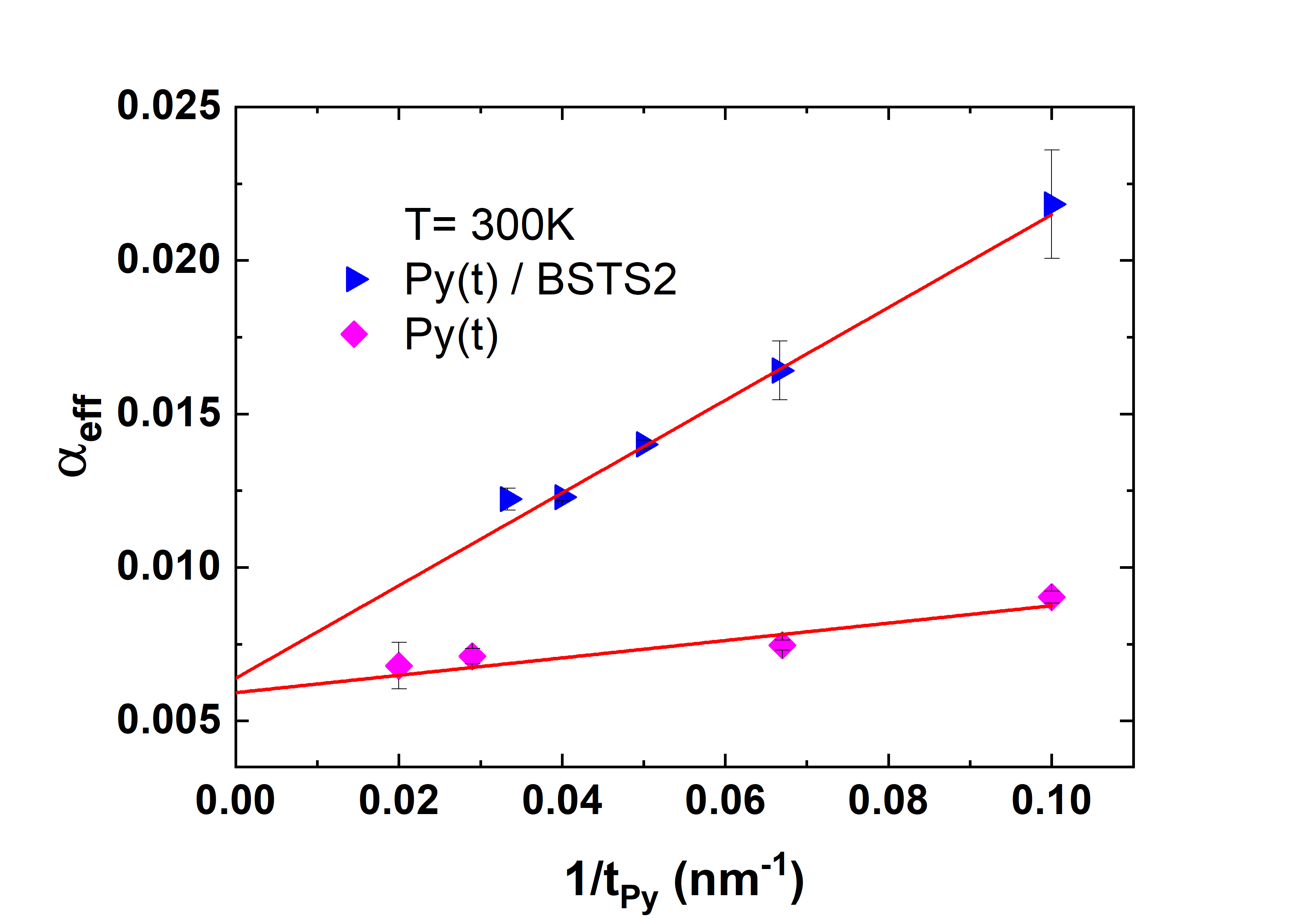}}
\subfigure[]{\includegraphics[width=7cm,height=5cm]{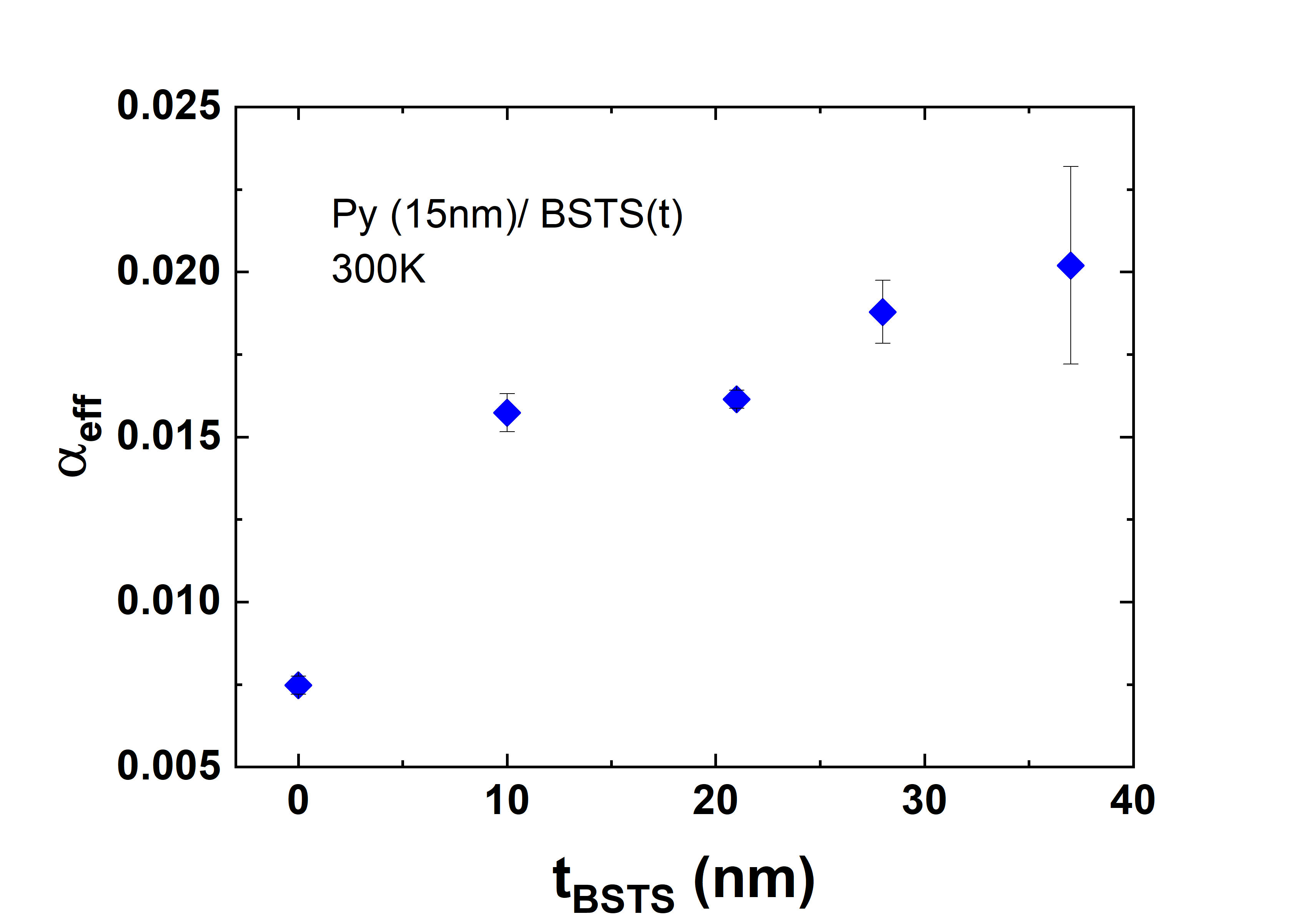}}
\caption{(a) $\alpha_{eff}$ variation with Py thickness for Py(t)/BSTS2 heterostructure at room temperature which fits in Eq.\ref{t_py}; (b) $\alpha_{eff}$ as a function of BSTS thickness for Py(15nm)/BSTS(t) heterostructure at room temperature.}
\label{Fig3}
\end{figure*}

For a given magnetic material at ferromagnetic resonance, the resonance field and frequency follow Kittel equation\cite{Kittel} given by,
\begin{equation}
f=\frac{\gamma}{2\pi}\sqrt{(H+H_k)(H+H_k+4\pi M_{eff})}
\label{Kittel}
\end{equation}
\begin{figure}
\centering
\subfigure[]{\includegraphics[width=7cm,height=5cm]{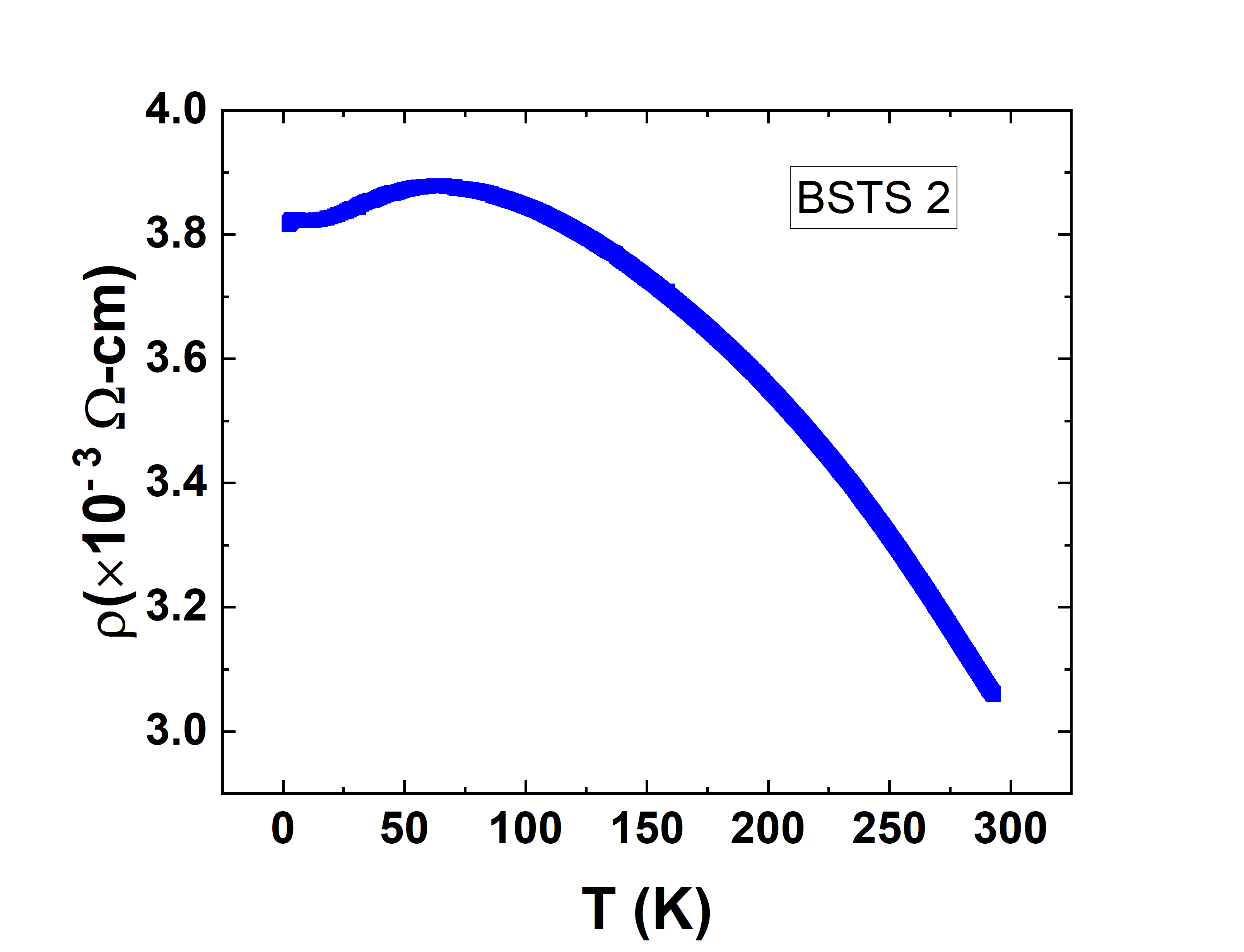}}
\caption{Temperature dependence of resistivity of the BSTS sample of thickness 21nm deposited on Si(111) substrate.}
\label{Fig4}
\end{figure}
\begin{figure*}
\centering
\subfigure[]{\includegraphics[width=7cm,height=5cm]{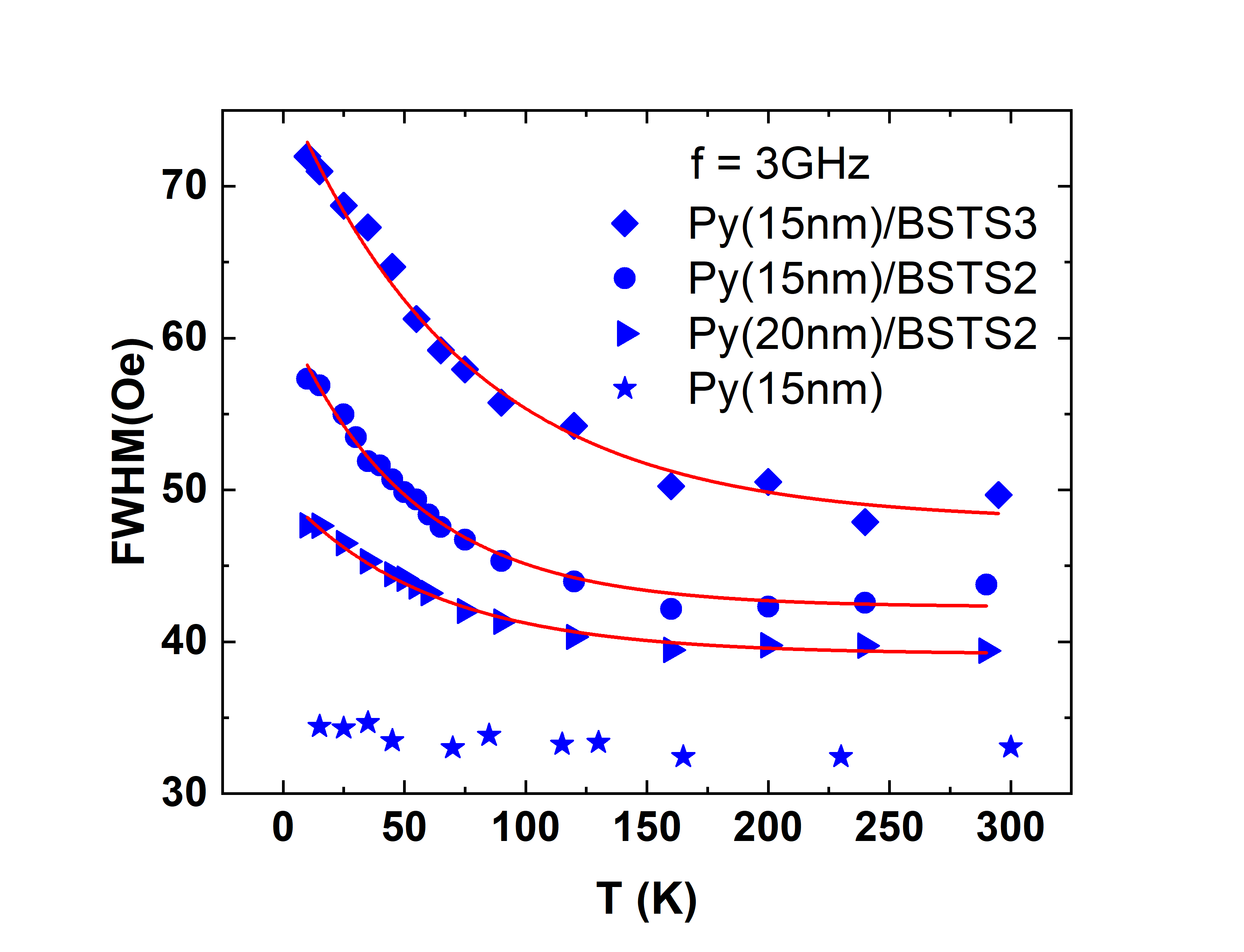}}
\subfigure[]{\includegraphics[width=7cm,height=5cm]{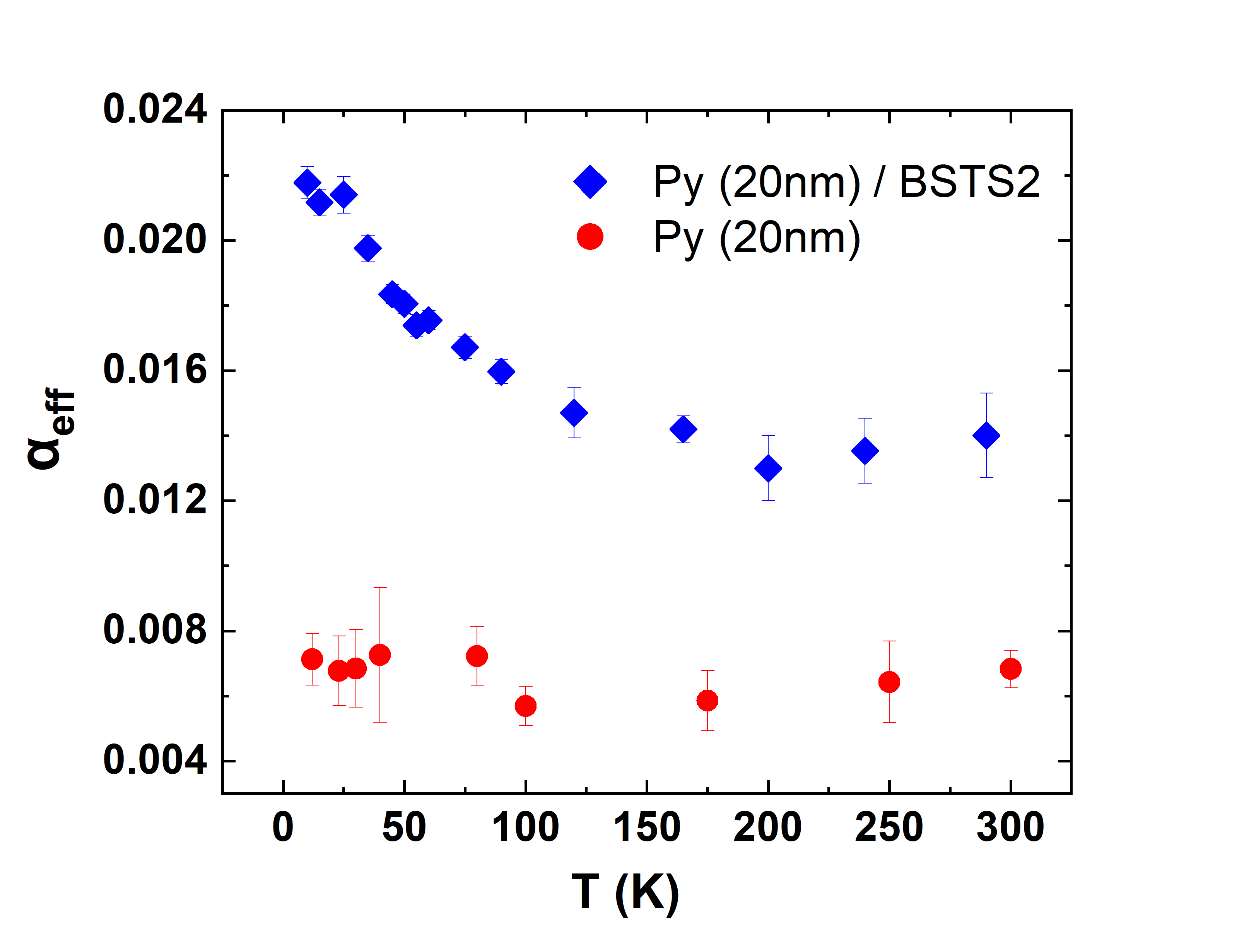}}
\caption{(a)Temperature dependence of the field linewidth ($\Delta H$) for different thickness combinations of Py/BSTS bilayer systems and for a bare Py thin film. The solid lines are the fits in the expression $exp(-T/T_{0})$; (b)Temperature dependence of effective damping coefficient, $\alpha_{eff}$ of Py(20nm)/BSTS2 and bare Py(20nm) film.}
\label{Fig5}
\end{figure*}

\begin{figure*}
\centering
\subfigure[]{\includegraphics[width=7cm,height=5cm]{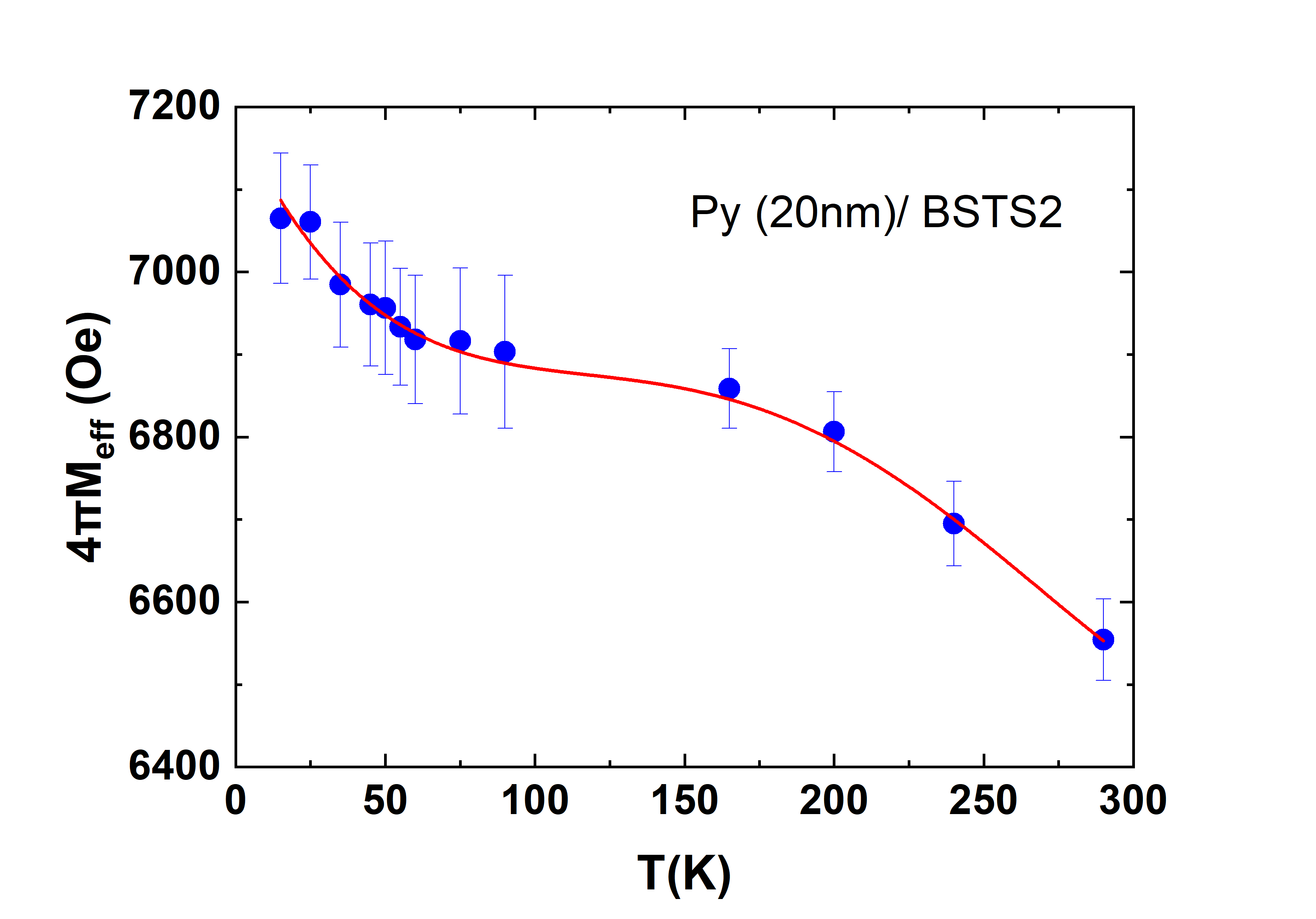}}
\subfigure[]{\includegraphics[width=7cm,height=5cm]{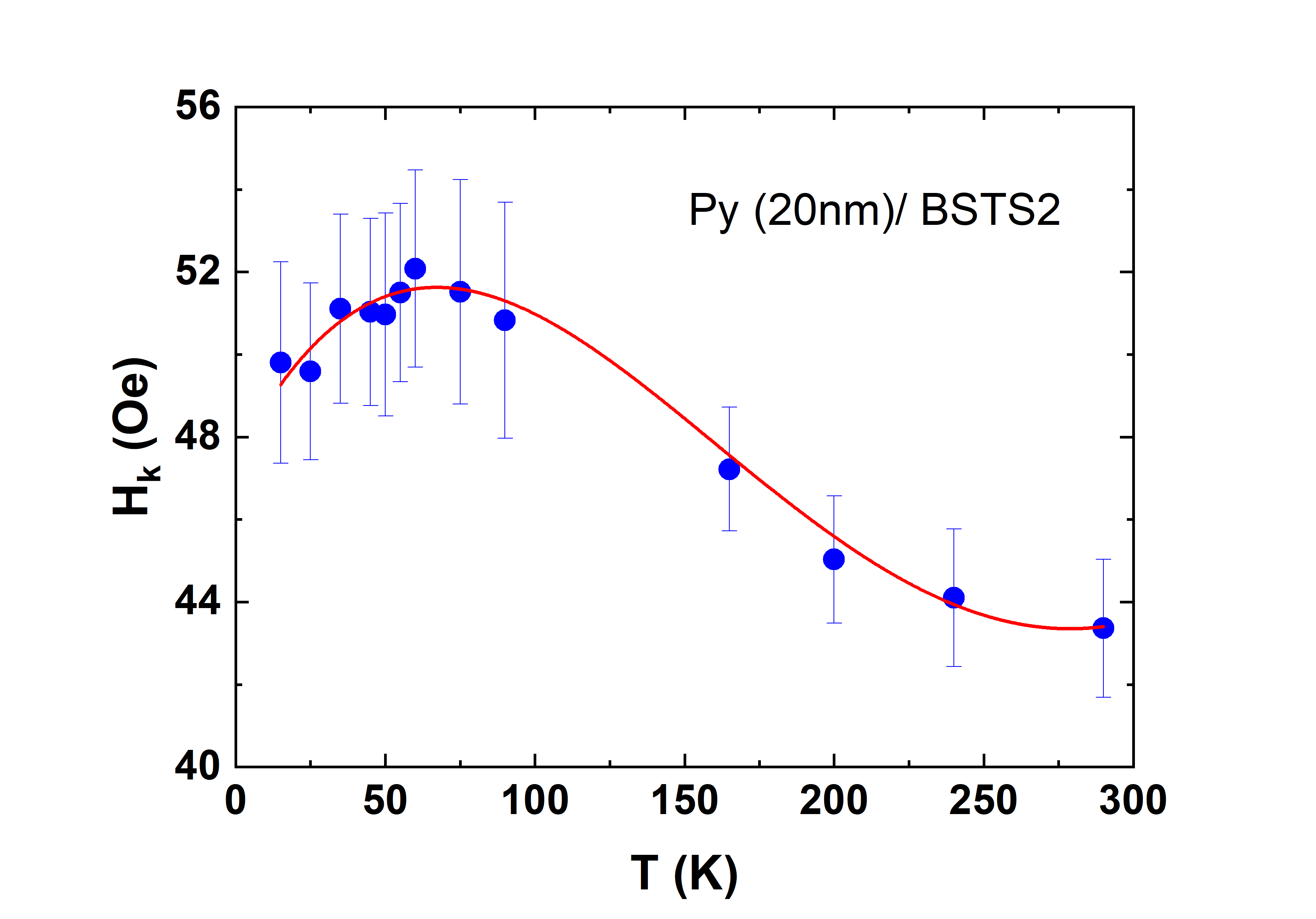}}
\caption{(a)Temperature dependence of effective magnetization of Py(20nm/BSTS2); (b)Temperature dependence of the anisotropy field of Py(20nm)/BSTS2.}
\label{Fig6}
\end{figure*}
where $H$, $H_k$, and $4\pi M_{eff}$ are the externally applied field,  magnetic anisotropy field, and effective magnetization respectively. We have obtained $H_k$ and $4\pi M_{eff}$ for different FM/TI bilayer systems by fitting the Kittel equation to the $f$ vs. $H$ curve as shown in fig.\ref{Fig2}b. The obtained $4\pi M_{eff}$ value contains saturation magnetization($4\pi M_s$) and other anisotropic contributions. We can evaluate $4\pi M_s$ value by analyzing the thickness dependent measurement of $4\pi M_{eff}$ of the FM layer. In the lower thickness region of the ferromagnetic thin films, $4\pi M_{eff}$ is inversely proportional to the film thickness and follows the equation\cite{M_s},
\begin{equation}
4\pi M_{eff}=4\pi M_s -\frac{2K_{s}}{M_s d}
\label{Ms}
\end{equation}
where $K_s$ is the surface anisotropy constant and $d$ is the thickness of the FM film.
The slope of the linear fit gives the anisotropy field contribution to $4\pi M_{eff}$ and the intercept gives the $4\pi M_s$ value as shown in fig.\ref{Fig2}c. The $4\pi M_{eff}$ does not depend on the thickness variation of BSTS at room temperature but $4\pi M_{eff}$ for Py(t) monolayer samples and for Py(t)/BSTS2 bilayer samples vary linearly with the inverse Py thickness as shown in Fig.\ref{Fig2}c. From the linear fitting (Eq.\ref{Ms}) of $4\pi M_{eff}$ vs. $1/t_{Py}$ data for the Py(t) and Py(t)/BSTS2 samples, we evaluated the saturation magnetization, $M_s$ of the Py/BSTS bilayer that has been decreased from that of the bare Py sample by an amount of 183$emu/cc^3$. It is a result of the loss of ferromagnetic order in the Permalloy layer. Due to the intermixing of the Py and BSTS at the interface, a magnetic dead layer could have formed at the interface which resulted in the reduction of saturation magnetization value as suggested by some previous studies \cite{dead3, dead2, dead1} also. The $K_{s}$ value has decreased from $0.092 \pm 0.008 erg/cm^2$ in bare Py film to $0.091 \pm 0.015 erg/cm^2$ in Py/BSTS2 bilayer. So interfacial anisotropy constant, $K_i$($=K_{s}^{Py/TI}-K_{s}^{Py}$) for the Py/BSTS2 sample is $-0.001 erg/cm^2$. From the negative value of $K_i$, we can ensure an in-plane magnetic anisotropy in the Py/BSTS interface at room temperature. A detailed discussion of magnetic anisotropy has been provided in the last section where the temperature variation of $H_k$ is discussed.\\
$\alpha_{eff}$ can be determined by analysing $\Delta H$ at different frequencies. $\Delta H$ contains both the intrinsic and extrinsic contributions to the damping. Linewidth due to intrinsic damping is directly proportional to the resonance frequency($f$) and follows the equation\cite{damping},
\begin{equation}
\Delta H=\Delta H_0+(\frac{2\pi\alpha_{eff}}{\gamma}) f
\label{Damping}
\end{equation}
where $\Delta H_{0}$ describes inhomogeneous linewidth broadening \cite{IL1,IL2} due to different extrinsic contributions like magnetic inhomogeneities \cite{IL3,IL4}, surface roughness, and defects in the sample. We have evaluated the $\alpha_{eff}$ values by fitting the $\Delta H$ vs $f$ curve for FM/TI bilayers as shown in fig.\ref{Fig2}a. This $\alpha_{eff}$ consists of Gilbert damping in the bulk ferromagnet($\alpha_{FM}$) and the enhanced damping($\alpha_{SP}$) resulting from spin pumping into the adjacent TI layer \cite{1a,1b,bra3}, $\alpha_{eff}=\alpha_{FM}+\alpha_{SP}$. The $\alpha_{FM}$ value for bare Py film of thickness 15nm was calculated to be 0.0074 and for the FM/TI bilayer system there has been significant enhancement in the $\alpha_{eff}$ value over the bare Py value due to spin pumping, $\alpha_{SP}$. In this heterostructure, $\alpha_{eff}$ increases gradually as the thickness of Py decreases both for Py(t) and Py(t)/BSTS2 samples as shown in fig.\ref{Fig3}a. From the linear fit of $\alpha_{eff}$ vs. $1/t_{Py}$ data we have obtained the spin-mixing coefficient, $g_{eff}^{\uparrow \downarrow}$ for the BSTS/Py interface to be $5.26\times 10^{18} \pm0.71 \times 10^{18} m^{-2}$ by using the equation\cite{FM/TI1,3b},
\begin{equation}
\alpha_{eff}-\alpha_{FM}=\frac{g\mu_B}{4\pi M_s t_{FM}} g_{eff}^{\uparrow \downarrow}
\label{t_py}
\end{equation}
where, $g$ and $\mu_{B}$ are the $g$-factor and Bohr magneton respectively. We have also calculated the spin current density($j_{s}^{0}$) for the FM/TI heterostructure using the $g_{eff}^{\uparrow\downarrow}$ value in the following equation\cite{3a,Sup3},
\begin{equation}
j_{s}^{0}=\frac{g_{eff}^{\uparrow\downarrow}\gamma^2 h_{m}^{2}\hbar[4\pi M_s \gamma+\sqrt{(4\pi M_s)^2 \gamma^2+4\omega^2}]}{8\pi\alpha^2[(4\pi M_s)^2\gamma^2+4\gamma^2]}
\label{J_s}
\end{equation}
where $\gamma$, $h_m$, $\hbar$, $\omega$, and $\alpha$ are the gyromagnetic ratio, microwave magnetic field, Planck's constant, Larmour precession frequency, and effective damping parameter respectively. Using Eq.\ref{J_s} the $j_s^0$ value for Py/BSTS2 was obtained to be $0.901\times 10^{-10}\pm 0.122\times10^{-10} Jm^{-2}$ in our experiment. The $g_{eff}^{\uparrow \downarrow}$ and $j_s^0$ values obtained from Py thickness-dependent study of $\alpha_{eff}$ are in a comparable range of the previously reported values for other combinations of ferromagnet and TI bilayer structures \cite{3b, FM/TI1, FM/TI2}. This gives evidence of successful spin injection into the BSTS layer from the Py layer due to spin pumping \cite{1a,1b,bra3, TI1}. We also report the TI thickness-dependent study of $\alpha_{eff}$ as shown in fig.\ref{Fig3}b. For bilayer structures of Py(15nm)/BSTS2(t) there is a sudden jump in the $\alpha_{eff}$ value from that of the bare FM film ($\alpha_{FM} = 0.0074$) because of spin pumping. Then with the thickness variation of TI layer in the range of 10nm to 37nm, $\alpha_{eff}$ increases slowly from 0.015 to 0.02. The TI thickness dependence of $\alpha_{eff}$ for Py(15nm)/BSTS(t) bilayer is almost linear which certainly can not be described by the conventional spin diffusion theory \cite{d2} for FM/NM proposed by Tserkovnyak \textit{et al.} \cite{d1}. For Py/BSTS heterostructure, $\alpha_{eff}$ vs. $t_{BSTS}$ study suggests an efficient spin-sink nature of the TI bulk with increasing thickness at room temperature \cite{ss}. From the room temperature study we certainly can not distinguish the TI surface state contribution from the TI bulk state contribution because growing a BSTS thin film with a perfectly insulating bulk state is still very challenging. Thus it was imperative to study the effect of topological surface state at low-temperature where bulk states of TI get suppressed and surface states of TI starts to dominate.\\

In this section, we have focused on low-temperature measurements specifically to understand the effect of topological surface states (TSS) on the magnetization relaxation of FM. At higher temperatures, a significant amount of bulk carriers are available to participate in the transport but with the reduction of phonon scattering, surface carriers dominate at a lower temperature.  From the resistivity vs. temperature data of BSTS2 in fig.\ref{Fig4}, we can see an insulating behavior of resistivity due to the enhanced insulating nature of the bulk state of TI at higher temperatures and a metallic behavior of resistivity below a certain temperature where the topological surface states dominate. We measured temperature variation of FMR linewidth ($\Delta H$), enhanced damping coefficient ($\alpha_{eff}$), anisotropy field ($H_k$) and effective magnetization ($4\pi M_{eff}$). For different thickness combinations of Py/BSTS bilayer, we obtained the $\Delta H$ variation with temperature. It increases exponentially with decreasing temperature that fits the expression, $exp(-T/T_{0})$ as shown in fig.\ref{Fig5}a. For bare Py(15nm) film, we can note that there is no significant variation in $\Delta H$ at low temperatures as can be seen from the curve at the bottom of fig.\ref{Fig5}a. To gain further understanding, the temperature variation of $\alpha_{eff}$ has also been studied for Py(20nm)/BSTS2 as shown in fig.\ref{Fig5}b and compared with $\alpha_{eff}$ for bare Py film. From the enhancement of $\alpha_{eff}$ value for Py(20nm)/ BSTS2 at room temperature we can ensure a successful spin injection due to the spin pumping effect. But the exponential increase of $\alpha_{eff}$ with decreasing temperature for the bilayer implies a huge increment in the amount of spin angular momentum transfer into the TI layer at lower temperatures. We attribute the origin of the exponential increase of $\alpha_{eff}$ and $\Delta H$ at lower temperatures to the spin chemical potential bias induced spin current into the surface state of TI as proposed by Abdulahad \textit{et al.} \cite{TI1}. The induced spin current into the TI surface state at lower temperatures corresponds to the rapid relaxation of magnetization precession of FM which is reflected in the exponential increase of $\Delta H$ and $\alpha_{eff}$ of the ferromagnet.\\

To further investigate the effect of TI surface state on the magnetization of FM, we studied the temperature variation of $4\pi M_{eff}$ and $H_k$ for Py(20nm)/BSTS2. In our previous study \cite{S3} with bare Py thin films, we have seen that $4\pi M_{eff}$ increases monotonically as saturation magnetization increases with lowering the temperature. But from fig.\ref{Fig6}a, we can see that the low-temperature dependence of $4\pi M_{eff}$ for Py/BSTS2 bilayer deviates from the single layer Py film [Supplementary fig.S11(a)]. This anomaly in $4\pi M_{eff}$ is related to the change of magnetic anisotropy energy of the system as well as the other effects like spin chemical potential induced current and exchange coupling between TSS and FM as mentioned by Abdulahad \textit{et al.} \cite{TI1}. In a previous section, we evaluated the interfacial magnetic anisotropy coefficient ($K_i=-0.001erg/cm^2$) to be in-plane of the interface of the Py/BSTS2 bilayer. The anisotropy field associated with the system anisotropy energy shows an interesting nature as we lower the temperature. We can see from fig.\ref{Fig6}b that $H_k$ increases initially with decreasing temperature until a certain value is reached and then the anisotropy field weakens against a further decrease in temperature. Thus we get a hump-like feature of $H_K$ for the same temperature region where $4\pi M_{eff}$ shows the anomaly and it is concomitant with the resistivity vs temperature behavior of the BSTS2 sample. The low-temperature behavior of $H_k$ and $4\pi M_{eff}$ can be justified by the argument proposed by Abdulahad \textit{et al.} \cite{TI1}. In their phenomenological model, they propose an existence of exchange interaction between the surface states of TI and local moments of the ferromagnetic layer. Several theoretical as well as experimental predictions confirm the existence of gapless topological surface states even after transition metal deposition on TI \cite{TTSS1, TTSS2}. These surface states can couple with the local moments of the FM through exchange interaction without any long-range ferromagnetic order. This exchange coupling acts perpendicular to the TI surface and weakens the in-plane anisotropy at lower temperatures where the surface states of TI dominate.


\section*{CONCLUSIONS}
In summary, we have carried out spin-pumping experiment in $BiSbTe_{1.5}Se_{1.5}$(TI)/ $Ni_{80}Fe_{20}$(FM) bilayer system. From the thickness-dependent measurements of FM/TI bilayers, we obtained the spin-transport parameters like damping coefficient due to spin-pumping, spin mixing conductance, and spin current density at room temperature. These results demonstrate a successful spin transfer from the FM layer to the TI layer due to spin-pumping. We have performed low-temperature measurements to specifically understand the surface state contribution of TI on the FM magnetization because the surface states of TI are more pronounced at lower temperatures. We have confirmed the suppression of the insulating bulk state of TI  at lower temperatures from the resistivity vs. temperature data of TI. In our low-temperature measurements of FMR linewidth and effective damping coefficient, we have witnessed an exponential increase in both parameters with the decrease in temperature. It suggests a spin chemical potential bias-induced spin current injection into the surface states of TI that gets enhanced at low temperatures \cite{TI1}. We have also studied temperature variations of the effective magnetization of the system. It showed a deviation from the bare Py film \cite{S3} in the temperature regime where TI surface states dominate. This deviation of effective magnetization results from the change in the anisotropy energy of the system. At room temperature, we evaluated the magnetic anisotropy energy coefficient which is found to be in-plane of the interface. This in-plane anisotropy weakens when conducting surface state of TI starts to dominate. It reflects from the hump-like feature in the magnetic anisotropy field vs. temperature data of the bilayer system. The decrease in in-plane magnetic anisotropy below a certain temperature can result from the exchange coupling between the surface states of TI and the local moments of the FM layer which act perpendicular to the interface \cite{TI1}. Combining the results of our low-temperature measurements we can conclude that there exists an exchange coupling between the TI surface state and FM which does not create any long-range ferromagnetic order in the TI and is unable to alter the overall spin texture of the TI surface state at the interface\cite{Sup1}. However, it affects the magnetization dynamics of the ferromagnetic metal quite significantly. These added features of enhancing the damping coefficients enables another fast control of magnetization dynamics in the FM layer.

\section*{ACKNOWLEDGEMENTS}
The authors sincerely acknowledge the Ministry of Education, Government of India and Science and Engineering Research Board (SERB) (grant no: EMR/2016/007950), and Department of Science and Technology (grant no. DST/ICPS/Quest/2019/22)
for financial support. S.P. acknowledges the Department of Science and Technology(DST)-INSPIRE fellowship India, S. A. acknowledges the Ministry of Education of the Government of India, S.M. acknowledges the Council Of Scientific and Industrial Research(CSIR), India, S.G.N and K.S acknowledges the University Grant Commission, India for research fellowship. The authors would like to thank Dr. Partha Mitra of the Department of Physics, Indian Institute of Science Education and Research Kolkata, for providing the lab facilities for sample deposition. The authors would like to acknowledge Prof. Anjan Barman and Mr. Pratap Kumar Pal of the Department of Physics, SN Bose National Centre for Basic Sciences for helping with thickness measurements using the XRR facility in their institute.


\end{document}